\documentstyle[11pt,newpasp,twoside,epsf]{article}
\begin{document}

\title{Fireballs, Flares and Flickering}
\author{K.\ J.\ Pearson}
\affil{Louisiana State University, Department of Physics and Astronomy, 
Nicholson Hall, Baton Rouge, LA~70803-4001, U.S.A.}
\author{Keith Horne}
\affil{School of Physics and Astronomy, University of St. Andrews, North Haugh,
 St. Andrews, Fife, KY16~9SS, U.K.}
\author{Warren Skidmore}
\affil{Department of Physics and Astronomy, University of California, Irvine, 
4129 Frederick Reines Hall, Irvine, CA~92697-4574, U.S.A.}

\begin{abstract} 
We review our understanding of the prototype ``Propeller'' system
AE~Aqr and we examine its flaring behaviour in detail. The flares are
thought to arise from collisions between high density regions in the 
material expelled from the system after interaction with the
rapidly rotating magnetosphere of the white dwarf. We show
calculations of the time-dependent 
emergent optical spectra from the resulting hot, expanding ball of gas
and derive values for the mass, lengthscale and temperature 
of the material involved. We see that the fits suggest that the
secondary star in this system has reduced metal abundances and 
that, counter-intuitively, the evolution of the fireballs is best modelled as 
isothermal. 
\end{abstract} 

\section{Introduction}
\label{sec:intro}

AE~Aqr is a long period CV ($P_{\rm orb}=9.88~{\rm h}$), implying that the
secondary is somewhat evolved.
Coherent oscillations in optical (Patterson 1979), ultraviolet 
(Eracleous et al.\ 1994), and X-ray (Patterson et al.\ 1980) lightcurves
reveal the white dwarf's 33s spin period. 
The oscillation has two unequal peaks per spin cycle,
consistent with broad hotspots on opposite sides of the white dwarf
$\sim 15^\circ$ above and below the equator (Eracleous et al.\ 1994).
These oscillations are strongest in the ultraviolet,
where their spectra show a blue continuum with broad Ly~$\alpha$ absorption
consistent with a white dwarf ($\log{g}\approx8$) atmosphere 
with $T\sim 3\times10^4~{\rm K}$ (Eracleous et al.\ 1994).
The simplest interpretation is accretion heating
near the poles of a magnetic dipole field tipped almost perpendicular
to the rotation axis.

An 11-year study of the optical oscillation period (de Jager et al.\ 1994) 
showed that the white dwarf is spinning down at an alarming rate.
Something extracts rotational energy from the white dwarf
at a rate $I\omega\dot{\omega} \sim 60 \nu L_\nu$ 
ie. some 60 times the luminosity of the system.
We now believe that to be a magnetic propeller. This model was
first proposed by Wynn, King, \& Horne (1995) and Eracleous \& Horne (1996)
and expanded upon in Wynn, King \& Horne (1997) with comparison of observed 
and modelled tomograms. Further work on the flaring region was reported
in Welsh, Horne \& Gomer (1998) and Horne (1999).

The gas stream emerging from the companion star through the L1 nozzle
encounters a rapidly spinning magnetosphere.
The rapid spin makes the effective ram pressure so high
that only a low-density fringe of material becomes threaded
onto field lines.
Most of the stream material remains diamagnetic, and is
dragged toward co-rotation with the magnetosphere.
As this occurs outside the co-rotation radius, this magnetic drag
propels material forward, boosting its velocity up to and beyond
escape velocity.
The material emerges from the magnetosphere and sails out of the
binary system.
This process efficiently extracts energy and angular momentum
from the white dwarf, transferring it via the long-range magnetic field
to the stream material, which is expelled from the system.
The ejected outflow consists of a broad equatorial fan of material
launched over a range of azimuths on the side away from the secondary..

The material stripped from the gas stream and threaded by the field
lines has a different fate, one which we believe gives rise to the
radio and X-ray emission.
This material co-rotates with the magnetosphere while accelerating
along field lines either toward or away from the white dwarf
under the influences of gravity and centrifugal forces.
The small fraction of the total mass transfer that leaks below the co-rotation 
radius at $\sim5~R_{\rm wd}$
accretes down field lines producing the surface hotspots 
responsible for the 33s oscillations.
Particles outside co-rotation remain trapped long enough to
accelerate up to relativistic energies through magnetic pumping,
eventually reaching a sufficient energy density to break away
from the magnetosphere (Kuijpers et al.\ 1997).
The resulting ejection of balls of relativistic magnetized plasma
is thought to give rise to the flaring radio emission 
(Bastian, Dulk \& Chanmugam 1998a,b).

In many studies the lightcurves exhibit dramatic flares,
with 1-10 minute rise and fall times 
(Patterson 1979; van Paradijs, Kraakman \& van Amerongen 1989; Bruch 1991;
Welsh, Horne \& Oke 1993).
The flares seem to come in clusters or avalanches of many super-imposed 
individual flares separated by quiet intervals of gradually
declining line and continuum emission 
(Eracleous \& Horne 1996; Patterson 1979).
These quiet and flaring states typically last a few hours.
Power spectra computed from the lightcurves have a power-law form,
with larger amplitudes on longer timescales.
Similar red noise power spectra are seen in 
active galaxies, X-ray binaries, and other cataclysmic variables,
and is therefore regarded as characteristic of accreting sources in general.
However, flickering in other cataclysmic variables
typically has an amplitude of 5-20\% (Bruch 1992), contrasting with
factors of several in AE~Aqr.
If the mechanism is the same, then it must be weaker or
dramatically diluted in other systems. The extreme behaviour in AE~Aqr may thus
allow us to probe the underlying physics involved where the resultant effects 
are clearest.

The optical and ultraviolet spectra of the AE~Aqr flares 
are not understood at present except in the most general terms.
The lines and continua rise and fall together, with little change
in the equivalent widths or ratios of the emission lines (Eracleous
\& Horne 1996). This suggests that the flares represent changes in the amount
of material involved more than changes in physical conditions.
Ultraviolet spectra from HST reveal a wide range of lines representing
a diverse mix of ionization states and densities.
Eracleous \& Horne (1996) concluded
that a large fraction of the line-emitting gas the density is in 
the range $n\sim10^{15}$--$10^{17}~{\rm m}^{-3}$ and that denser 
regions likely also exist. The C{\sc \small IV} emission is unusually weak, 
suggesting non-solar abundances consistent with the idea that the
secondary is evolved. Latest results of ultraviolet observations and
abundance analysis of the secondary are presented by Mouchet (this volume)
and Bonnet-Bidaud (this volume).

What mechanism triggers these dramatic optical and ultraviolet flares?  
Clues come from multi-wavelength co-variability and 
orbital kinematics.
Simultaneous VLA and optical observations show that the radio 
flux variations occur on similar timescales but are not correlated
with the optical and ultraviolet flares,
which therefore require a different mechanism (Abada-Simon et al.\ 1995).
It has been proposed that the flares represent modulations
of the accretion rate onto the white dwarf, so that they should be
correlated with X-ray variability.
Some correlation was found, but the correlation is not high.
However, HST observations discard this model, because
the ultraviolet oscillation amplitude is unmoved by
transitions between the quiet and flaring states (Eracleous \& Horne 1996).
This disconnects the origins of the oscillations and flares,
and the oscillations arise from accretion onto the white dwarf,
so the flares must arise elsewhere.

Further clues come from emission line kinematics.
The emission line profiles may be roughly described as
broad Gaussians with widths $\sim1000$~km~s$^{-1}$,
though they often exhibit kinks and sometimes multiple peaks.
Detailed study of the Balmer lines (Welsh et al.\ 1998) indicates that the
new light appearing during a flare can have emission lines
shifted from the line centroid and somewhat narrower,
$\sim300$~km~s$^{-1}$.
Individual flares therefore occupy only a subset of the entire
emission-line region.

The emission line centroid velocities vary sinusoidally
with orbital phase, with semi-amplitudes $\sim200$~km~s$^{-1}$
and maximum redshift near phase $\sim$0.8 (Welsh et al.\ 1998).
These unusual orbital kinematics are shared by both 
ultraviolet and optical emission lines.
The implication is that the flares arise from gas
moving with a $\sim200$~km~s$^{-1}$ velocity vector
that rotates with the binary and points away the observer at phase 0.8.
This is hard to understand in the standard model
of a cataclysmic variable star, though many (eg. SW Sex systems)
show similar anomalous emission-line kinematics (Thorstensen et al.\ 1991) 
suggesting a common underlying mechanism. 
In particular, Kepler velocities in an AE~Aqr accretion disc 
would be $>600$~km~s$^{-1}$ (though we believe no disc to be present.)
The gas stream has a similar direction but its velocity is
$\sim1000$~km~s$^{-1}$.
A success of the magnetic propeller model is its ability to
account for the anomalous emission-line kinematics.
The correct velocity amplitude and direction occurs
in the exit fan just outside the Roche lobe of the white dwarf.
But the question remains of why the flares are ignited here, several hours
after the gas slips silently through the magnetosphere.

\begin{figure}
\plotfiddle{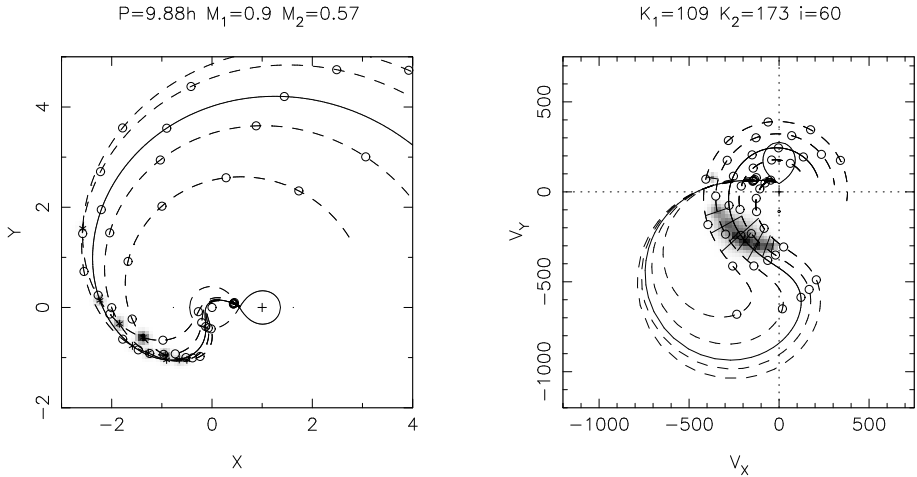}{2.5in}{270}{50}{50}{-200}{250}
\caption{Trajectories of diamagnetic blobs passing though the AE~Aquarii 
system with the corresponding Doppler tomogram. The open circles mark the 
time of flight in units of $0.1~P_{\rm orb}$. Asterisks and grey patches
mark the locations where lower density blobs overtake and collide with more 
compact blobs, producing fireballs. This occurs in the lower left quadrant
at the location corresponding to the observed emission lines.}
\label{fig:trajdop}
\end{figure}

The key insight which solved this puzzle was the realization
that the magnetic propeller acts as a blob sorter.
More compact, denser, diamagnetic blobs are less affected by magnetic drag.
They punch deeper into the magnetosphere and
emerge at a larger azimuth with a smaller terminal velocity.
These compact blobs can therefore be overtaken by `fluffier' blobs
ejected with a larger terminal velocity in the same direction,
having left the companion star somewhat later
but having spent less time in the magnetosphere (Wynn et al.\ 1997).
The result is a collision between two gas blobs, which can give rise
to shocks and flares.
Calculations of the trajectories of magnetically propelled diamagnetic
blobs with different drag coefficients indicate that they
cross in an arc-shaped region of the exit stream, 
in just the right place to account for the orbital kinematics of
the emission lines (Welsh et al.\ 1998; Horne 1999). Figure~\ref{fig:trajdop} 
shows how these trajectories map to a locus
of points in the lower left quadrant of the Doppler map that is 
otherwise difficult to populate.

There remains the problem of developing a quantitative understanding
of the unusual emission-line spectra resulting from the collision.

\section{Fireball Models}
\label{sec:models}

\subsection{Typical Parameters}
\label{sec:estimate}

We can derive rough estimates for the physical parameters associated with
the colliding blobs by 
using the typical rise time $t\sim300~{\rm s}$ and
an observed optical flux $f_{\nu}\sim50$~mJy with a  
closing velocity $V\sim300~{\rm km~s}^{-1}$ for the blobs, together with
a distance of 100pc. A summary of typical values is provided in 
Table~\ref{tab:scaleval}.

\begin{table}
\begin{center}
\begin{tabular}{rrl} \hline
Quantity & \multicolumn{2}{c}{Value} \\ \hline 
observed flux $f_{\nu}$ & $50$ & mJy\\
closing velocity $V$ & $300$ & $\rm{km~s}^{-1}$\\
flare risetime $t_{\rm rise}$ & 300 & s \\
mass transfer rate $\dot{M}$ & $10^{14}$ & ${\rm kg~s}^{-1}$ \\
fireball mass $M$ & $3\times10^{16}$ & kg \\
pre-collision lengthscale $a$ & $9\times10^{7}$ & m \\
initial temperature $T$ & $10^{6}$ & K \\
total energy $E$ & $3\times10^{26}$ & J \\
typical density $\rho$ & $3\times10^{-8}$ & ${\rm kg~m}^{-3}$ \\
number density $n$ & $3\times10^{19}$ & ${\rm m}^{-3}$ \\
column density $N$ & $3\times10^{27}$ & ${\rm m}^{-2}$ \\ \hline 
\end{tabular}
\end{center}
\caption{Estimates of typical values for flare quantities.}
\label{tab:scaleval}
\end{table}

\subsection{Expansion}

We envision an expanding fireball emerging from the aftermath of
a collision between two masses $m_1+m_2=M$.
With nothing to hold it back, the hot ball of gas expands
at the initial sound speed, launching a fireball. We adopt a uniform, 
spherically symmetric, Hubble-like
expansion $V=Hr_{0}=Hr/\beta$, in which the Eulerian radial coordinate $r$
 of a gas element is given 
in terms of its initial position $r_0$ and time $t$ by
\begin{equation}
r(r_{0},t)=r_{0}+v(r_{0})t=r_{0}+Hr_{0}t=r_{0}(1+Ht)\equiv r_{0}\beta.
\label{expfac}
\end{equation}
This defines an expansion factor $\beta$ which we can use as a dimensionless 
time parameter. The `Hubble' constant is set by the initial conditions
$H a_{\rm i} \approx v(a_{\rm i}) \approx c_{\rm s,i}$.
If we can determine parameters at some time $t=0$ ie. $\beta=1$ for the
lengthscale ($a_0$), temperature ($T_0$) and mass ($M$), we can derive the time
evolution as follows. We adopt a Gaussian density profile
\begin{equation}
\rho (r,t) = \rho_{0} \beta^{-3} e^{-\eta^{2}}
\label{denprof}
\end{equation}
where $\rho_{0}$ is the central density at $t=0$ and 
$\eta\equiv\frac{r_{0}}{a_{0}}=\frac{r}{a}$
is a dimensionless radius coordinate: the radius $r$ scaled to the lengthscale
$a=\beta a_{0}$. This Gaussian density is motivated by the Gaussian 
shapes of observed, Doppler-broadened, line profiles. 

\subsection{Cooling}

We consider two cooling schemes: adiabatic and isothermal.
The adiabatic fireball cools purely as a result of its expansion
and corresponds to a situation where the radiative and 
recombination cooling rates are negligible. In contrast, the isothermal model 
maintains a fixed temperature throughout its evolution. A truly isothermal 
fireball would require a finely balanced energy source to 
counteract the expansion cooling. However, it may be an appropriate 
approximation to a 
situation where a photospheric region dominates the emission and presents a  
fixed effective temperature to the observer as a result of the stong 
dependence of opacity on temperature. For adiabatic expansion we can show that
 an initially uniform temperature distribution remains uniform and so,
\begin{equation}
	T(r,t) = T_0 \beta^{-3(\gamma-1)}.
\label{temptdep}
\end{equation}
With $\gamma=5/3$ for a monatomic gas, $T\propto \beta^{-2}$, and so
the sound speed $c_{\rm s} \propto T^{1/2} \propto \beta^{-1}$.
The sound crossing time $a/c_{\rm s} \propto \beta^2$.
and so the fireball becomes almost immediately supersonic.

\subsection{Ionization Structure}

In the LTE approximation, the density and temperature
determine the ionization state of the gas at each point in space and time
through the solution of a network of Saha equations.
Atomic level populations are similarly determined 
through Boltzmann factors and partition functions.

Once we have determined the evolution of $T$ and $\rho$ with time, 
we can follow the evolution of the ionization structure. 
At a given time, the various ionic species form onion layers of increasing
ionization on moving outwards from the fireball centre. The
adiabatic temperature evolution causes the ionization states to alter rapidly
throughout the structures when the fireball passes the appropriate
critical temperatures. Isothermal evolutions show the gentler dependance of
ionization on density. The spatial structure remains roughly constant and only
evolves slowly as the density drops.

\section{Radiative Transfer}
\label{sec:radtrans}

The radiative transfer equation has a formal solution
\begin{equation}
	I = \int S e^{-\tau} d\tau , \label{radtrans}
\end{equation}
where $I$ is the intensity of the emerging radiation,
$S$ is the source function, and $\tau$ is the optical depth
measured along the line of sight from the observer.
This integral sums contributions $S d\tau$ to the radiation intensity,
attenuating each by the factor $e^{-\tau}$ because 
it has to pass through optical depth $\tau$ to reach the
observer.

Since we have assumed LTE, the source function is the Planck function,
and opacities both for lines and continuum are also
known once the velocity, temperature, and density profiles
and element abundances are specified.
The integral can therefore be evaluated numerically, either in this form or 
more  quickly by using Sobolev resonant surface approximations. 

The above line integral gives the intensity
$I(y)$ for lines of sight with different impact parameters $y$.
We let $x$ measure distance from the fireball centre toward the
observer, and $y$ the distance perpendicular to the line of sight.
The fireball flux, obtained by summing intensities
weighted by the solid angles
of annuli on the sky, is then
\begin{equation}
f(\lambda) = \int_0^\infty I(\lambda,y) \frac{2 \pi y}{ d^2 } dy
\end{equation}
where $d$ is the source distance.

\section{Comparison with Observations}
\label{sec:results}

\subsection{Observed Flare in AE Aqr}

\begin{figure}
\plotfiddle{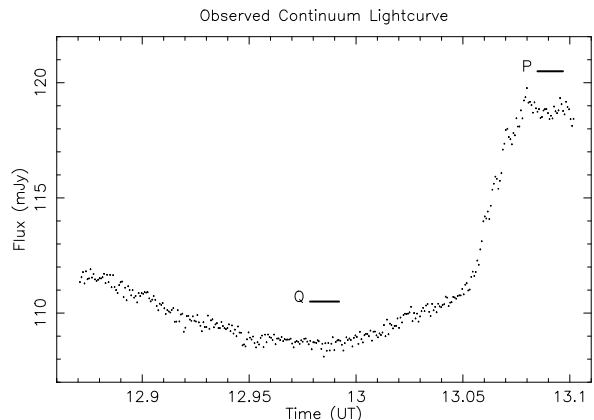}{2.0in}{270}{30}{30}{-100}{150}
\caption{Observed lightcurve reported in Skidmore et al.\ (2003) showing
the decline of an old flare and start of a small new flare.}
\protect\label{fig:datlight}
\end{figure}

To test the fireball models, we compared them to 
high time-resolution optical spectra of an AE Aqr 
flare taken with the Keck telescope (Skidmore et al.\ 2003). The lightcurve 
formed from these data is
plotted in Fig.~\ref{fig:datlight}. Although the dataset only lasts
around 13 minutes it shows the end of one flare and beginning of 
another. The major features of the flaring observed in the 
system are apparent in the lightcurve.

\begin{figure}
\plotfiddle{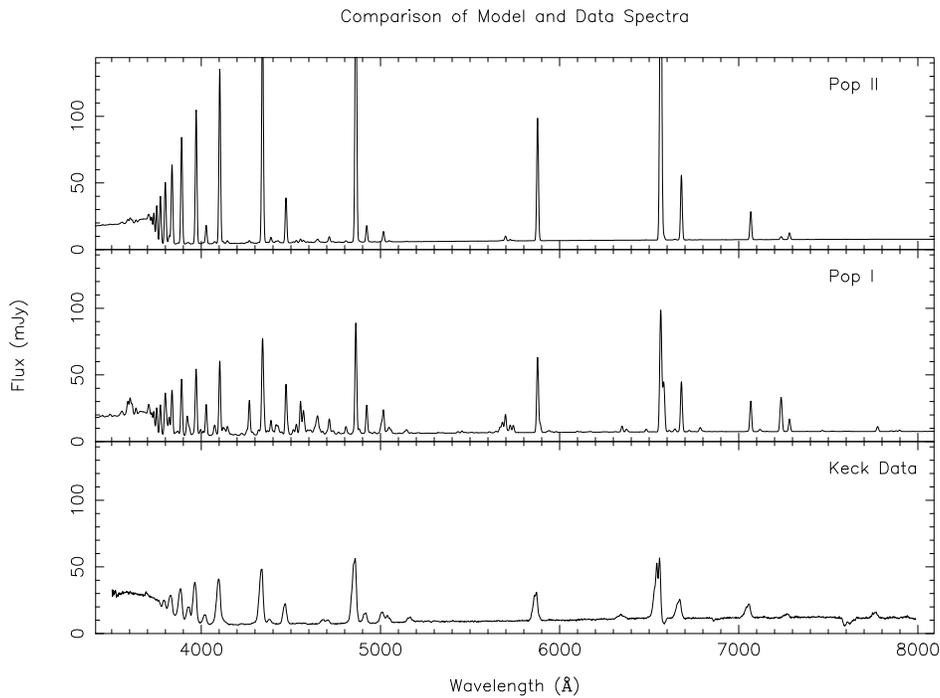}{3.5in}{270}{50}{50}{-175}{260}
\caption{Population I and II models compared to the observed data.}
\label{fig:popcomp}
\end{figure}

We extracted the spectrum of the second flare by subtracting
the quiescent spectrum (Q) from the spectrum at the top of the new 
flare (P). The resultant optical spectrum is plotted in the bottom panel of
Fig.~\ref{fig:popcomp}.

\subsection{Spectral Fit at Peak of Flare}

To estimate the fireball parameters $M$, $T$ and $a$ from the observed 
optical spectra we began with consideration of the continuum flux. We 
considered both Population I and Population II abundances. Population I 
abundances were set using the solar abundances presented by 
(D\"{a}ppen 2000).
For population II composition, we decreased from 0.085 to 0.028 
the ratio of the helium to hydrogen number densities and reduced the ratio 
for each metal abundance to 5\% of solar (Bowers \& Deeming 1984). 
In both cases we used an amoeba algorithm 
(Press et al.\ 1986) to optimize the fit by reducing $\chi^{2}$ to a minimum 
for 6 optical continuum fluxes. The best fit parameters
$M$, $T$ and $a$ and related secondary quantities are listed in 
Table~\ref{fitdata}. 

Both compositions arrive at a model fit with remarkably similar calculated
fluxes which have a slightly steeper slope to the Paschen continuum than
the observations. The fits also produce a large Balmer jump but appears to
have difficulty in creating one quite as strong as that observed. 
The central densities implied by both sets of
parameters are consistent with the typical values for the gas stream in
section~\ref{sec:estimate} and with the temperature range and emitting area
of Beskrovnaya et al.\ (1996).

\begin{table}
\begin{center}
\begin{tabular}{rcc}
\hline
Quantity & Pop I & Pop II \\ \hline
$M~(10^{16}~{\rm kg})$ & $3.7$ & $6.8$ \\ 
$a_{0}~(10^{7}~{\rm m})$ & $5.1$ & $9.6$ \\
$T_{0}$~(K) & $17~000$  & $18~000$ \\
$v(a_{0})~({\rm km~s}^{-1})$ & $170$ & $170$\\
$1/H$~(s) & $300$ & $560$ \\
$\rho_{0}~(10^{-8}~{\rm kg~m}^{-3})$ & $5.1$ & $1.4$\\ \hline
\end{tabular}
\end{center}
\caption{Parameters used for the model spectra and useful secondary 
quantities.}
\protect\label{fitdata}
\end{table}

We derived the expansion rate from consideration of the H$\beta$
and nearby HeI equivalent widths. In a fireball model, increasing the 
expansion velocity spreads out the line opacity and thereby increases the 
equivalent widths of the emission lines as long as they remain optically thick.
The H$\beta$ region was chosen 
because the observed H$\alpha$ line
shows a complex structure. We blurred the model spectra 
with a Gaussian of $9.8~\mbox{\AA}$ FWHM and assumed a distance of $102$~pc 
(Friedjung 1997). The H$\beta$ and HeI line equivalent widths suggest 
expansion velocities of $240~\mbox{km}~\mbox{s}^{-1}$ and
$100~\mbox{km}~\mbox{s}^{-1}$ respectively at $a_0$. 
Consequently, we adopted an expansion velocity at $a_0$ of 
$170~\mbox{km}~\mbox{s}^{-1}$ for the simulations.

The optical spectra produced with both compositions are compared to the 
observed mean spectrum at the flare peak in Fig.~\ref{fig:popcomp}. 
We see that the parameters, derived from the peak continuum fluxes alone, 
allow us to derive optical spectra with integrated line fluxes and ratios 
comparable to the 
observations. Both models have line widths narrower than the observations,
and hence peak line fluxes greater than those observed, which may result from
additional unidentified instrumental blurring. The Population I models
exhibit a 
flat Balmer decrement of saturated Balmer lines that is also apparent
in the data. Comparing the lines that are present, however, particularly in
the $4000-5000$~\AA\ range, the Population II model appears to produce a much 
better fit: suppressing the metal
lines which do not appear in the observed spectra. 

\subsection{Lightcurves and Spectral Evolution}

The timescale for 
the evolution of the fireball implied by the observed lightcurve can be 
reproduced by the parameters derived solely from fitting to the 
peak spectrum. The evolutionary timescale $a_{0}/v(a_{0})=1/H$ is 
approximately $300$~s and $560$~s for Population I and II parameters
respectively (Table~\ref{fitdata}). The encouraging result that these values 
are similar to the 
observed timescale suggests that our fireball models are on the right track.

\begin{figure}
\plotfiddle{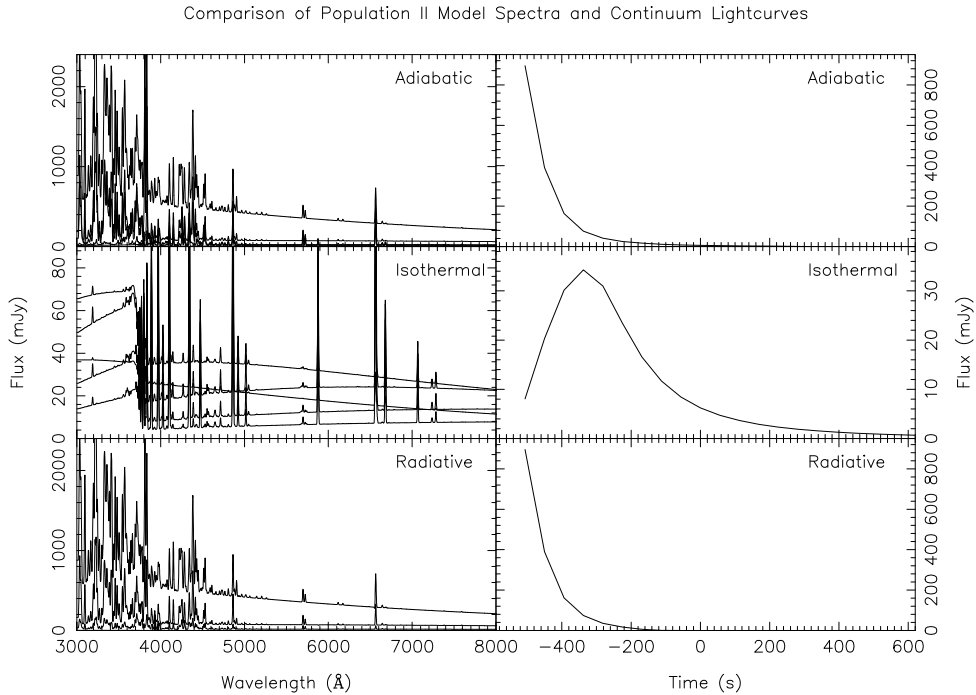}{3.75in}{270}{50}{50}{-190}{275}
\caption{Population II spectra at $t=-450,-337,-225,-113,0~\mbox{s}$, 
($\beta=0.2,0.4,0.6,0.8,1.0$) for the three cooling laws alongside their 
respective continuum lightcurves at $5~350$~\AA.}
\label{fig:popiimods}
\end{figure}

Fig.~\ref{fig:popiimods}  shows the lightcurves and spectral 
evolution for Population II abundances derived for isothermal and
adiabatic evolutions (`radiative' models will be discussed later).
In each case, the model fireball is constrained to evolve through the state
fitted to the observations at the peak of the flare.

The isothermal models produces a lightcurve that rises and
falls, as observed.  
The isothermal fireball spectra, both on the rise and on
the fall of the flare, exhibit strong Balmer emission
lines and a Balmer jump in emission.
Thus the isothermal fireball model provides a plausible fit not
only to the peak spectrum but also to the time evolution of
the AE Aqr flare.

The adiabatic model fails to reproduce the
observed lightcurve or spectra.  The adiabatic model
matches the observed spectrum at one time,
but is too hot at early times and too cool at late times.
As a result the lightcurve declines monotonically from
early times rather than rising to a peak and then falling,
and the spectra have lines that do not match the observed spectra.
This disappointing performance is perhaps not too surprising,
since the temperature in this model falls through a wide range
while the observations suggest that the temperature is
always around $10^4$~K. To explain the rise phase of the observed lightcurve,
for an adiabatic evolution, we must interpret it as 
resulting from the collision of the initial gas blobs. 

\section{Discussion}
\label{sec:discuss}
The spectra produced with Population II composition reproduce the
optical observations well. The fitted parameters are consistent with both the 
expected conditions in the mass transfer stream and with the lower limits
on the density implied by the ultraviolet observations 
($n\sim10^{15}~\mbox{m}^{-3}$).
The presence of high ionization and semi-forbidden
lines in the ultraviolet data may be understood qualitatively in terms of 
their 
formation in the low
density outer fringes of the fireball and the wide variety of ionization 
states by the wide range of densities in the expanding fireball structure.
For $\beta=1$, mean molcular weight $\mu=0.53$ and Population II parameters, 
the density
of $10^{17}~\mbox{m}^{-3}$, important for the ultraviolet semi-forbidden lines,
occurs at $\eta\approx2.3$. We can show that
this lies outside the limit for LTE behaviour (Pearson, Horne \& Skidmore 
2003).
Quantitative fits for the fringe and ultraviolet behaviour, therefore, 
remain to be addressed in the non-LTE regime. 

The behaviour of the lightcurve for an isothermal model is much
more similar to the observed lightcurve than the adiabatic model, giving a 
peak 
close to the observed time without the need to invoke the collision process
to create a rising phase.
Clearly though, an expanding gas ball would be expected to cool both from
radiative and adiabatic expansion effects. 

In short, we find that isothermal fireballs reproduce the observations rather 
well, 
whereas adiabatically cooling fireballs fail miserably! How can the expanding 
gas ball present a nearly constant temperature to the observer?

\subsection{Thermostat}

We examine two possible mechanisms which may operate to maintain the apparent 
fireball temperature in the $1$--$2\times10^4~\mbox{K}$ region. The first 
relies on the fact that both free-free and bound-free opacity decrease with 
temperature and thus opacity peaks just above the temperature at which 
hydrogen begins to recombine.
In this model a hot core region is surrounded by a cooler blanket of material.
Radiating at the hotter temperature, the core would rapidly lose its thermal 
energy through radiative cooling. 
However, we can envision a situation where the material just outside the core 
has reached the high opacity regime and is absorbing significant amounts of 
energy from the core's radiation field. This would help to counteract the 
effect of adiabatic cooling in this shell and may maintain an effectively 
isothermal blanket around the core in the following way. If the temperature in
the blanket were to rise, the opacity would decrease, more energy would 
escape and the
blanket would cool again. Similarly, until recombination, cooling would
result in higher opacity and therefore greater heating of the blanket region.
The core region would thus only be cooling through adiabatic expansion.
For such a thermostatic mechanism to work, the optical depth through the
 blanket region would need to be high which would be consistent with it being 
the photosphere as seen by an outside observer. 

Simple models have been attempted to mimic the effect of a photosphere
shrinking due to radiative cooling but these have been unsuccesful in 
reproducing the observations (Pearson et al.\ 2003). This occurs 
because the cooling front occurs well outside the photosphere. The
thin shell of cooling material at around $10^{4}~\mbox{K}$ does not provide 
sufficient opacity to 
mimic the isothermal behaviour and we therefore ``see'' into the 
adiabatically-cooling core.
The radiative model thus shows very similar behaviour to the adiabatic models. 

Realistic simulations of such a model require a time-dependent
radiation hydrodynamical treatment of 
the radiation field and its heating effect; a more sophisticated method 
than the one so far employed.

\subsection{External Photoionization}

\begin{figure}
\plotfiddle{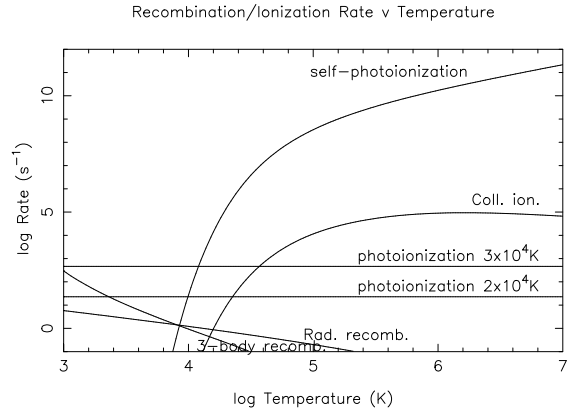}{2in}{270}{30}{30}{-100}{150}
\caption{Ionization and recombination rates for a hydrogen atom or ion 
respectively with $n_{\rm e}=3\times10^{18}~\mbox{m}^{-3}$.}
\label{fig:ionrate}
\end{figure}

Alternatively, the ionization of the fireball might be held at a temperature
typical of the  $1$--$2\times10^4~\mbox{K}$ range through photoionization by 
the  white dwarf. In
Fig.~\ref{fig:ionrate}, we plot the hydrogen ionization and recombination 
rates as a function of temperature for a typical electron number density 
$n_{\rm e}=3\times10^{18}~{\rm m}^{-3}$ using routines for photoionization 
from Verner et al.\ (1996), radiative recombination (Verner \& Ferland 1996),
three-body recombination (Cota 1987) and collisional ionization by Verner 
using the 
results of Voronov (1997). We calculate a conservative overestimate for
the self-photoionization of the fireball assuming the sky is half filled by
blackbody radiation at the given fireball temperature. Also shown is the
photoionization due to a blackbody with the same radius and distance 
as the white dwarf. The general white dwarf temperature is uncertain but,
as mentioned earlier, there is evidence to suggest that the 
hot spots have a temperature of around $3\times10^{4}~\mbox{K}$. We can see 
that the flux from the white dwarf overtakes that from the fireball itself as 
the dominant ionization mechanism at $\sim10^{4}~\mbox{K}$. We have assumed
here a typical white dwarf to fireball distance equal to 
that of the L1 point from the white dwarf (although in a different direction). 
The recombination 
rates for this plot have been calculated assuming $n_{\rm e}$ is a constant 
and so for temperatures below about
$10^{4}~\mbox{K}$ are conservative overestimates of the LTE recombination 
rate that will occur in our fireball. In spite of this,
the white dwarf photoionization at $3\times10^{4}~\mbox{K}$ 
comfortably exceeds the 
recombination rates down to at least $10^{3}~\mbox{K}$ and, hence, we would 
expect the fireball to remain almost completely ionized in this case.
This suggests that the white dwarf radiation field may be the cause of the 
fireball appearing isothermal when we consider the
lightcurve behaviour and consistent line strengths and ratios. 
Ionization by an external black body at a different temperature will
clearly cause our fireball to no longer be in LTE and require a
non-LTE model to follow the fireball behaviour accurately.

\section{Summary}
\label{sec:summary}

We have shown for AE~Aqr how the observed flare spectrum and evolution is 
reproducible with an isothermal fireball at $T\approx18~000$~K with 
Population II 
abundances but not when adiabatic cooling is incorporated. We suspect that 
the cause of the apparently isothermal nature is a combination of two 
mechanisms. First, a nearly isothermal
photosphere which is self-regulated by the temperature dependance of 
the continuum opacity and the hydrogen recombination front and second, 
particularly at late times, by the ionizing effect of the white dwarf
radiation field. 

The observed photosphere in the isothermal models is 
initially advected outward with the flow before
the decreasing density causes the opacity to drop and the photosphere to 
shrink to zero size. This gives rise to a lightcurve that rises and then falls.
Emission lines and edges arise because the photospheric radius is larger at 
wavelengths with higher opacity. Surfaces of constant Doppler shift are
perpendicular to the line of sight and thus Gaussian density profiles give rise
to Gaussian velocity profiles.

In a purely LTE fireball we expect high ionization states to
occur in the outer regions as a result of the lower density. 
In reality, of course, once the LTE boundary is crossed,
the ionization will drop as the net recombination rate gradually ekes away at 
the ions and we make a transition towards coronal equilibrium. 
In principle, we can follow the non-LTE
evolution of the gas through this outer region and beyond the ion-electron
equilibrium radius to the non-equilibrium ionization states present in the 
outer fringe. The higher ionization would then result from the 
higher temperature of
the fireball when the region in question crossed the LTE boundary. In practice,
we expect that this correction may have little effect on the optical spectra
because the emission is dominated by the higher density inner regions of the
fireball. This effect may well become important, however, when we
consider the ultraviolet spectra where high ionization lines are present; 
and which are not reproduced by the current models.

Improved modelling of these fireballs offers us the chance to probe the
chemical composition of the secondary star in AE Aqr in a complimentary regime
to those presented by Bonnet-Bidaud (this volume). Preliminary
results from SS Cyg encourage us that the fireball models will
be applicable more generally to the flickering observed in most CV
systems. In such `disc flickering' systems, the fireballs may
be launched by local heating from magnetic reconnection events rather than from
blob-blob collisions. Similar spectral features are present in the flickering 
data (see Skidmore et al. in this volume) with similar temporal evolution to 
the AE~Aqr flares. Initial fitting using the same procedure as for the 
AE Aqr observations give good results that suggest that a common mechanism may 
indeed be at work. 

\subsection*{Acknowledgements}

KJP would like to thank PPARC, NSF grant 115-30-5177 and the IAU whose 
financial support made attendance at the colloquium possible.


\begin{references}
\reference Abada-Simon, M., Bastian, T.\ S., Horne, K.\ D., Robinson, E.\ L., 
    Bookbinder, J.\ A. 1995, in Buckley, D.\ A.\ H., Warner, B., eds.,
    Proc. Cape Workshop on Magnetic Cataclysmic Variables, ASP Conf. Series,
    355 
\reference    Bastian, T.\ S., Dulk, G.\ A., Chanmugam, G.
    1988a, \apj, 324, 431 
\reference    Bastian, T.\ S., Dulk, G.\ A., Chanmugam, G.
    1988b, \apj, 330, 518 
\reference    Beskrovnaya, N., Ikhsanov, N., Bruch, A., Shakhovskoy, N. 
    1996, \aap, 307, 840
\reference    Bowers, R.\ L., Deeming, T. 1984, Astrophysics I Stars, 
    Jones and Bartlett, Boston 
\reference Bruch, A. 1991, \aap, 251, 59
\reference    Bruch, A. 1992, \aap, 266, 237 
\reference    Cota, S.\ A. 1987, PhD thesis, Ohio State Univ. 
\reference    D\"{a}ppen, W.
    2000, in Cox, A.\ N., Allen's Astrophysical Quantities, Chapter 3, 
    Springer-Verlag, New York
\reference   Eracleous, M., Horne, K.\ D., Robinson, E.\ L., Zhang, E.-H., 
Marsh, T.\ R., Wood, J.\ H. 1994, \apj, 433, 313 
\reference   Eracleous, M., Horne, K.\ D. 1996, \apj, 471, 427  
\reference    Friedjung, M. 1997, NewA, 2, 319
\reference    Horne, K.\ D.
    1999, in Hellier, K., Mukai, K., eds, 
    Annapolis Workshop on Magnetic Cataclymic Variables, ASP Conf. Series,
    157, 357
\reference   de Jager, O.\ C., Meintjes, P.\ J., O'Donoghue, D., 
Robinson, E.\ L. 1994, \mnras, 267, 577  
\reference   Kingdon, J.\ B., Ferland, G.\ H. 1996, \apjs, 106, 205  
\reference   Kuijpers, J., Fletcher, L., Abada-Simon, M., Horne, K.\ D.,
   Raadu, M. A., Ramsay, G., Steeghs, D. 1997, \aap, 322, 242
\reference van~Paradijs, J., Kraakman, H., van~Amerongen, S. 1989, \aaps, 
    79, 205
\reference Patterson, J. 1979, \apj, 234, 978
\reference   Patterson, J., Branch, D., Chincarini, G., Robinson, E.\ L.
    1980, \apj, 240, L133
\reference  
    Pearson, K.\ J., Horne, K.\ D., Skidmore, W. 2003, \mnras, 338, 1067
\reference  Press, W.\ H., Teukolsky, S.\ A., Vetterling, W.\ T., 
   Flannery, B.\ P. 1986, Numerical Recipes in Fortran, Cambridge Univ. Press
\reference  Skidmore, W., O'Brien, K., Horne, K.\ D., 
    Gomer, R., Oke, J.\ B., Pearson, K.\ J. 2003, \mnras, 338, 1057
\reference  Thorstensen, J.\ R., Ringwald, F.\ A., Wada, R.\ A., 
  Schmidt, G.\ A.,  Norsworthy, J.\ E. 1991, \aj, 102, 272
\reference  Verner, D.\ A., Ferland, G.\ J. 1996, \apjs, 103, 467
\reference  Verner, D.\ A., Ferland, G.\ J., Korista, K.\ T., Yakovlev, D. G.
   1996, \apj, 465, 487
\reference  Voronov, G.\ S. 1997, Atomic Data and Nuclear Data Tables, 65, 1
\reference Welsh, W.\ F., Horne, K., Oke, B. 1993, \apj, 406, 229
\reference  Welsh, W.\ F., Horne, K.\ D., Gomer, R. 1998, \mnras, 298, 285
\reference  Wynn, G.\ A., King, A.\ R., Horne, K.\ D. 1995, 
    in Buckley, D.\ A.\ H.,
    Warner, B., eds., Cape Workshop on Magnetic Cataclysmic Variables, 
    ASP Conf. Series, 85, 196, Astron. Soc. Pacific.,  San Francisco
\reference  Wynn, G.\ A., King, A.\ R., Horne, K.\ D. 1997, \mnras, 286, 436
\end{references}
\end{document}